\documentclass[11pt]{article}
\usepackage{verbatim,url,enumerate,color,paralist}
\usepackage{amsmath,amsfonts}
\usepackage{epsfig,amssymb,amstext,xspace,theorem}
\usepackage{algorithm}
\usepackage{algorithmicx,algpseudocode}
\usepackage{fullpage}

\newcommand{\qed}{\hspace*{\fill}$\Box$}
\newtheorem{theorem}{Theorem}[section]
\newtheorem{lemma}[theorem]{Lemma}
\newtheorem{corollary}[theorem]{Corollary}

\newtheorem{remark}[theorem]{Remark}

\newenvironment{proof}[1][Proof. ]{\noindent {\bf #1 }}{\qed}

\newcommand{\Xomit}[1]{}

\begin{document}

\title{An LP-based $\frac{3}{2}$-approximation algorithm for the graphic $s$-$t$ path TSP}
\author{
Zhihan Gao\thanks{
        (z9gao@uwaterloo.ca)
	Dept.\ of Comb.\ \& Opt.,
        University of Waterloo, Waterloo, Ontario N2L3G1, Canada.
	}
}

\date{}


\maketitle


\begin{abstract}
We design a new LP-based algorithm for the graphic $s$-$t$ path Traveling Salesman Problem (TSP), which achieves the best approximation factor of $1.5$. The algorithm is based on the idea of narrow cuts due to An, Kleinberg, and Shmoys. It partly answers an open question of Seb\H{o}.
\end{abstract}


\section{Introduction \label{sec:intro}}

The metric Traveling Salesman Problem (TSP) is one of the most well-known problems in the area of combinatorial optimization. For the metric TSP, Christofides \cite{christofides76} presented an algorithm that
achieves an approximation guarantee of $\frac{3}{2}$. Hoogeveen \cite{hoogeveen91} extended the algorithm to the metric $s$-$t$ path TSP, and proved an approximation guarantee of $\frac{5}{3}$. This had been the best approximation factor for decades until the recent paper \cite[An, Kleinberg, and Shmoys]{AKS12} improved on the $\frac{5}{3}$ approximation guarantee and presented an algorithm that achieves an approximation guarantee of $\frac{1+\sqrt{5}}{2}\approx1.61803$. Most recently, \cite[Seb\H{o}]{sebo12} further improved the approximation factor to $1.6$.

For the graphic $s$-$t$ path TSP, a special case of the metric $s$-$t$ path TSP, \cite[An and Shmoys]{AS11} provided a sightly improved performance guarantee of $(\frac{5}{3}-\epsilon)$. The paper \cite[M{\"o}mke and Svensson]{MS11} gave a $1.586$-approximation algorithm for the graphic $s$-$t$ path TSP. \cite[Mucha]{Mucha12} improved the analysis of \cite{MS11} and obtained a $\frac{19}{12}+\epsilon\approx1.58333+\epsilon$ approximation guarantee for any $\epsilon > 0$ for the graphic $s$-$t$ path TSP. Recently, \cite[Seb\H{o} and Vygen]{SV12} gave the first $1.5$-approximation algorithm for the graphic $s$-$t$ path TSP. Their algorithm and its analysis are sophisticated, and are based on ear decomposition. The algorithm applies both local and global optimization to the ears.

In this paper, we present a new $1.5$-approximation algorithm for the graphic $s$-$t$ path TSP. Compared with the algorithm from \cite[Seb\H{o} and Vygen]{SV12}, our algorithm and its analysis are much simpler. The notion of narrow cuts for $s$-$t$ path TSP was introduced by \cite[An, Kleinberg, and Shmoys]{AKS12}. Our algorithm is based on this idea. In \cite{sebo12}, Seb\H{o} posed an open question on applying the "Best of Many Christofides" algorithm in \cite{AKS12} to achieve the best approximation guarantees known for the graphic special cases of TSP and its variants. Although our algorithm seems different from the "Best of Many Christofides" algorithm, they share the idea of narrow cuts. From this point of view, our algorithm answers this question partly for the graphic $s$-$t$ path TSP. The key point of our algorithm is to find a minimal spanning tree that intersects every narrow cut in an odd number of edges. Such a tree guarantees that the minimum size of the edges fixing the wrong degree vertices is at most half of the optimal value of the linear programming relaxation. Finally, the union of the spanning tree and the fixing edges gives us the $1.5$-approximation guarantee.
\section{Preliminaries \label{sec:prelims}}
Let $G=(V, E)$ be a connected graph with unit cost on each edge. Let $s,t$ be two given vertices in $G$.
Consider the metric completion $(G^{\prime}, c^{\prime})$ of $G$
where $c^{\prime}$ is the cost function on each edge $e=(u, v)$ in $G^{\prime}$ such that $c^{\prime}_e$
is the minimal cost of any $u$-$v$ path in $G$. The \textit{graphic $s$-$t$ path TSP} is to find a minimum cost Hamiltonian path from $s$ to $t$ in $G^{\prime}$ with edge costs $c^{\prime}$. Denote the cost of this path by $OPT(G)$.

For any vertex subset $\phi \subsetneq S \subsetneq V$, we define $\delta_G(S)=\{(u, v)\in E: u\in S, v\notin S\}$. If there is no ambiguity, we use $\delta(S)$ for short. In particular, if $S=\{v\}$, we use $\delta(v)$ instead of $\delta(\{v\})$. Let $\mathcal{W}=\{W_1, W_2, \ldots, W_l\}$ be a partition of vertex set $V$. Define $\delta(\mathcal{W})=\cup_{1\leq i\leq l}\delta(W_i)$. Let $x\in \mathbb{R}^{E}$. For any $F\subseteq E$, we define $x(F)=\sum_{e\in F} x_e$. Let $2G$ be the graph obtained from $G$ by doubling every edge of $G$. The graphic $s$-$t$ path TSP of $G$ is equivalent to finding a minimum-size trail in $2G$ from $s$ to $t$ visiting every vertex at least once (multiple visits are allowed). That is to find a minimum-size connected spanning subgraph of $2G$ with $\{s,t\}$ as the odd-degree vertex set. The following linear program (LP) on the original graph $G$ is a relaxation of the graphic $s$-$t$ path TSP:

$
\begin{array}{rrcll}
\hbox{({\bf L.P.1})}\quad  {\rm minimize}: & \sum_{e \in E} x_e & & \\
{\rm subject~to}: & x(\delta(\mathcal{W})) & \geq & |\mathcal{W}|-1 & \forall \mbox{ partition } \mathcal{W} \mbox{ of } V \\
& x(\delta(S)) & \geq & 2 & \forall \emptyset \subsetneq S \subsetneq V, |S \cap \{s, t\}| \mbox{ even } \\
& 2 \geq & x_e & \geq   0 & \forall e \in E
\end{array}
$

Let $x^*$ be an optimal solution of (L.P.1). Note that (L.P.1) can be solved in polynomial time via the ellipsoid method \cite{GLS81}. We know that $\sum_{e \in E} x^{*}_e \leq OPT(G)$. Let $S\subseteq V$. If $|S\cap \{s, t\}|=1$, we call $S$ an \textit{$s$-$t$ cut}. Furthermore, if $x^*(\delta(S))<2$, we call $S$ a \textit{narrow cut}.

\begin{lemma}\cite[Lemma 1]{AKS12}\label{lemma:narrowcut}
Let $S_1, S_2\subseteq V$ be two distinct narrow cuts such that $s\in S_1$ and $s\in S_2$. Then $S_1\subsetneq S_2$ or $S_2\subsetneq S_1$.
\end{lemma}

Hence, we know that the set of narrow cuts containing $s$ forms a nested family. Let $S_1, S_2, \ldots, S_k$ be all the narrow cuts containing $s$ such that $s\in S_1 \subsetneq S_2 \subsetneq S_3 \cdots \subsetneq S_k\subsetneq V$. Define $L_i=S_i\backslash S_{i-1}$ for $i=1, 2, \ldots, k, k+1$ where $S_0=\phi$ and $S_{k+1}=V$. Note that each $L_i$ is nonempty and $\cup_{1\leq i\leq k+1}L_i=V$.

Let $T$ be a nonempty subset of $V$ with $|T|$ even. For $F\subseteq E$, if the set of odd-degree vertices of graph $(V, F)$ is $T$, then we call $F$ a \textit{$T$-join}. Note that if $G$ is connected, then a $T$-join always exists. For any $S\subseteq V$, if $|S\cap T|$ is odd, we call it \textit{$T$-odd cut}. The following LP formulates the problem of finding a $T$-join of minimum size:

$
\begin{array}{rrcll}
\hbox{({\bf L.P.2})}\quad  {\rm minimize} : & \sum_{e \in E} x_e & & \\
{\rm subject~to} : & x(\delta(S)) & \geq & 1 & \forall \mbox{ $T$-odd } S \\
& x_e & \geq & 0 & \forall e \in E
\end{array}
$

\begin{lemma}\cite{EJ01}\label{lem:Tjoin}
The optimal value of (L.P.2) is the same as the minimum size of a $T$-join.
\end{lemma}

Let $F\subseteq E$. For any $v\in V$,  we call $v$ a wrong degree vertex with respect to $F$ if
\begin{align}
 |\delta(v)\cap F| \mbox{ is }
 \begin{cases}
  even \hspace{0.3cm} \text{ if } v\in \{s, t\} \\
  odd  \hspace{0.3cm} \text{ if } v\notin \{s, t\}.
 \end{cases}
\end{align}
We use the next lemma through the rest of the paper.
\begin{lemma}\cite[Lemma 1]{CFG12}\label{lem:cutparity}
Let $G=(V,E)$ be a graph, let $s$, $t$ be two vertices of $G$, let $F$ be a set of edges of $G$,
and let $T$ be the set of wrong-degree vertices with respect to $F$. Then,
for any $S\subseteq{V}$, if $S$ is $T$-odd and also satisfies $|S\cap \{s,t\}|=1$,
then $|\delta(S)\cap F|$ is even.
\end{lemma}

\section{LP-based $\frac{3}{2}$-approximation algorithm}

In this section, we give an LP-based $\frac{3}{2}$-approximation algorithm for $s$-$t$ path TSP. Before stating the algorithm, we need some lemmas.

\begin{lemma}\label{findNC}
There is a polynomial-time combinatorial algorithm to find all narrow cuts $S_1, S_2, \ldots, S_k$.
\end{lemma}
\begin{proof}
Compute the Gomory-Hu tree for the terminal vertex set $V$ with respect to the capacity $x^*$ (See \cite[Section 3.5.2]{CombOptBook}). After that, for each edge of the $s$-$t$ path in the Gomory-Hu tree, check the corresponding cut. We claim that each such cut with $x^*$ capacity less than $2$ is a narrow cut, and there are no other narrow cuts. The correctness of this claim follows from the following observation: For any $u\in L_i$, $v\in L_{i+1}$, the narrow cut $S_i$ is the unique minimum $u$-$v$ cut, and furthermore, $S_i$ is also a $s$-$t$ cut.
\end{proof}

Let $H$ be the support graph of $x^*$. For any $L\subseteq V(H)$, the subgraph of $H$ induced by $L$ is denoted by $H(L)$.
\begin{lemma}
For $1\leq p\leq q\leq k+1$, $H(\cup_{p\leq i\leq q} L_i)$ is connected.
\end{lemma}
\begin{proof} Consider the graph $H$ which is the support graph of $x^*$. Note that $x^*(\delta_H(S))=x^*(\delta_G(S))$ for any $\phi \subset S\subset V$. In this proof, the notation refers to $H$, e.g., $\delta(S)$ means $\delta_H(S)$. Let $L=\cup_{p\leq i\leq q} L_i$. We divide the proof into several cases:\\
\noindent \textbf{Case 1:} $p=1$ and $q=k+1$, i.e., $H=H(L)$. The partition cut constraint in (L.P.1) implies that $H$ is connected. \\
\noindent \textbf{Case 2:} $p=1$ and $q<k+1$. Suppose $H(L)$ is not connected. Then, there exist two nonempty vertex sets $U_1$ and $U_2$ such that $U_1, U_2$ is a partition of $L$ and there exists no edge between $U_1$ and $U_2$ in $H$. Without loss of generality, we can assume that $s\in U_1$. By the constraints of (L.P.1), we have $x^*(\delta(U_1))\geq 1$ and $x^*(\delta(U_2))\geq 2$. However, $L=S_{q}$ is a narrow cut, which implies $x^*(\delta(L))<2$. Note that $\delta(U_1)\cap \delta(U_2)=\phi$ and $\delta(L)=\delta(U_1)\cup \delta(U_2)$. So, $2>x^*(\delta(L))=x^*(\delta(U_1))+x^*(\delta(U_2))\geq 1+2=3$. This is a contradiction.\\
\noindent \textbf{Case 3:} $p>1$ and $q=k+1$. By the symmetry of $s$ and $t$, it is the same as Case 2.\\
\noindent \textbf{Case 4:} $p>1$ and $q<k+1$. Suppose $H(L)$ is not connected. Then, similarly there exist two nonempty vertex sets $U_1$ and $U_2$ such that $U_1, U_2$ is a partition of $L$ and there exists no edge between $U_1$ and $U_2$ in $H$. In this case, by the constraints of (L.P.1), we have $x^*(\delta(U_1))\geq 2$ and $x^*(\delta(U_2))\geq 2$. Let $Y_1=\cup_{1\leq i\leq q} L_i$ and $Y_2=\cup_{p\leq i\leq k+1} L_i$. Note that $Y_1$ and $Y_2$ are two narrow cuts. Also, $\delta(U_1)\cup \delta(U_2) \subseteq \delta(Y_1)\cup \delta(Y_2)$. Note that $\delta(U_1)\cap \delta(U_2)=\phi$ by the definition of $U_1$ and $U_2$. Thus, $4>x^*(\delta(Y_1))+x^*(\delta(Y_2))\geq x^*(\delta(Y_1)\cup \delta(Y_2))\geq x^*(\delta(U_1)\cup \delta(U_2))=x^*(\delta(U_1))+x^*(\delta(U_2))\geq 2+2=4$. This is a contradiction.
\end{proof}
\begin{corollary}\label{conTr}
For every $1\leq i\leq k+1$, $H(L_i)$ is connected, and moreover there exists an edge connecting $L_i$ and $L_{i+1}$ in $H$.
\end{corollary}

\begin{algorithm}
Step 1. Find an optimal solution $x^*$ of (L.P.1) and construct the support graph $H$ of $x^*$.\\

Step 2. Find the narrow cuts $S_1, S_2, \ldots, S_k$ containing $s$, and get the corresponding sets $L_1, L_2, \ldots, L_{k+1}$ (recall: $L_i=S_i\backslash S_{i-1}$ where $S_0=\phi$ and $S_{k+1}=V$). If no narrow cuts exist, take $J$ as a spanning tree in $G$ and go to Step 6.\\

Step 3. For $1\leq i\leq k+1$, find a spanning tree $J_i$ on $H(L_i)$.\\

Step 4. Take an edge $e_i$ from $H$ connecting $L_i$ to $L_{i+1}$ for $1\leq i\leq k$. Let $E_b=\cup_{1\leq i \leq k} \{e_i\}$.\\

Step 5. Construct a spanning tree $J=(\cup_{1\leq i\leq k+1} J_i) \cup E_b$.\\

Step 6. Let $T$ be the wrong degree vertex set of $J$. Find the minimum size $T$-join $F$ in $G$.\\

Step 7. Output $J\dot{\cup} F$ (disjoint union of edge sets in $2G$).
\caption{LP-based approximation for the graphic $s$-$t$ path TSP}
\end{algorithm}
Lemma \ref{findNC} provides a polynomial algorithm for Step 2, and Corollary \ref{conTr} guarantees that Step 3 and Step 4 are feasible. Thus, the LP-based algorithm runs in polynomial time.
\begin{lemma} For $F$ in the LP-based algorithm, we have
\[
|F|\leq \frac{1}{2}\sum_{e\in E}x^*_e.
\]
\end{lemma}
\begin{proof}
 Firstly, we claim $x^*(\delta(S))\geq 2$ for every $T$-odd cut where $T$ is the wrong degree vertex set of $J$ in the LP-based algorithm. Let $S$ be a $T$-odd cut. There are two cases to be considered.

\noindent \textbf{Case 1:} $S$ is not an $s$-$t$ cut. Then, by the constraint of (L.P.1), we have $x^*(\delta(S))\geq 2$ \\
\noindent \textbf{Case 2:} $S$ is an $s$-$t$ cut. If there exist no narrow cuts, then clearly $x^*(\delta(S))\geq 2$. Otherwise, for any narrow cut $S^{\prime}$, we have $|J\cap \delta(S^{\prime})|=1$ by Step 4 of the algorithm. However, by Lemma \ref{lem:cutparity}, we have $|J\cap \delta(S)|$ is even. This means $S$ is not a narrow cut. Thus, $x^*(\delta(S))\geq 2$. \\
By the claim, we know $\frac{1}{2}x^*(\delta(S))\geq 1$ for every $T$-odd cut $S$. This implies $\frac{1}{2}x^*$ is a feasible solution of (L.P.2). By Lemma \ref{lem:Tjoin}, we have $|F|\leq \frac{1}{2}\sum_{e\in E}x^*_e$. This completes the proof.
\end{proof}
\begin{remark}
In fact, if we can find a spanning tree $J$ such that $|J\cap \delta(S)|$ is odd for each narrow cut $S$, then we can find an edge set $F$ to correct the wrong degree vertices in $J$ such that $|F| \leq \frac{1}{2}\sum_{e\in E}x^*_e$. This also holds for the (general) metric $s$-$t$ path TSP, i.e., for metric costs, if we can find such a spanning tree $J$, then the minimum cost of the edges fixing the wrong degree vertices in $J$ is at most half of the cost of the optimal solution of LP.
\end{remark}
\begin{theorem}\label{mainThm}
The LP-based algorithm is a $\frac{3}{2}$-approximation for the graphic $s$-$t$ path TSP.
\end{theorem}

\begin{proof}Note that $J$ is a spanning tree of $G$. We consider $J$ as an edge set. So, $|J|=|V|-1\leq \sum_{e\in E}x^*_e\leq OPT(G)$. Also note that $|F|\leq \frac{1}{2}\sum_{e\in E}x^*_e\leq \frac{1}{2}OPT(G)$. Since $J\dot{\cup} F$ is a connected spanning subgraph of $2G$ with $\{s, t\}$ as the odd-degree vertex set, this gives a $s$-$t$ Hamiltonian path on the metric completion of $G$ with cost at most $|J|+|F|$. Therefore, the LP-based algorithm is a $\frac{3}{2}$-approximation algorithm.
\end{proof}

\begin{remark}
By the proof of Theorem \ref{mainThm}, we can obtain an upper bound $\frac{3}{2}$ for the integrality ratio of the (L.P.1). Furthermore, this also implies that the integrality ratio of the path-variant Held-Karp relaxation (See \cite{AKS12}) is at most $\frac{3}{2}$ when restricted to graphic metric. Note that \cite[Figure 1(b)]{AKS12} presented an example with graphic metric to show the lower bound  $\frac{3}{2}$ for the integrality ratio of the path-variant Held-Karp relaxation. Hence, from this point of view, our algorithm achieves the best possible approximation guarantee that the LP-based algorithms can get for the graphic $s$-$t$ path TSP.
\end{remark}


\medskip
\noindent
{\bf Acknowledgements}.
The author is grateful to Joseph Cheriyan for stimulating discussions and indispensable help, and to Zachary Friggstad, Laura Sanit\`{a} and Chaitanya Swamy for their useful comments.

\bibliographystyle{abbrv}

\end{document}